\documentclass[prb,final,twocolumn,showpacs,preprintnumbers,amsmath,amssymb]{revtex4}

\usepackage[]{graphicx}

\begin{document}

\title{\textbf{Hole redistribution in Sr$_{14-x}$Ca$_{x}$Cu$_{24}$O$_{41}$ ($x$=0, 12) spin ladder
compounds: $^{63}$Cu and $^{17}$O NMR study under pressure.}}

\date{\today}
\author{ Y. Piskunov$^{1,2}$, D. J\'{e}rome$^{1}$,
P. Auban-Senzier$^{1}$, P. Wzietek$^{1}$ and A. Yakubovsky$^{3}$}
\affiliation {$^{1}${ Laboratoire de Physique des Solides (UMR
8502)}, Universit\'{e} Paris-Sud, 91405, Orsay, France\\
$^{2}${ Institute of Metal Physics }, Ekaterinburg 620219, Russia\\
$^{3}${ Russian Scientific Center ''Kurchatov Institute''},
Moscow, Russia}
\date{\today}

\begin{abstract}

We report the results of a $^{63}$Cu and $^{17}$O  NMR study of
the nuclear quadrupole interaction tensor,
$^{17,63}\nu_{Q,\alpha}$, in the hole doped spin ladder system
Sr$_{14-x}$Ca$_{x}$Cu$_{24}$O$_{41}$ ($x$=0 and 12) performed
under ambient  and high pressures. NMR data show that the hole
density in the Cu$_{2}$O$_{3}$ ladder layer grows with
temperature, Ca content and an applied pressure. We have derived
the  hole occupation of  Cu 3d and O 2p orbitals at the different
ion sites in the Cu$_{2}$O$_{3}$ ladder as a function of the
temperature, Ca substitution and pressure. We also suggest that
the most important role of high pressure for the stabilization of
a superconducting ground state in  Ca-rich two-leg ladders is an
increase of the hole concentration in the conducting
Cu$_{2}$O$_{3}$ planes. We have obtained an estimate  of 0.10 hole
per Cu1 for the hole concentration at low temperature in Ca12
under 32 kbar when this compound undergoes a superconducting transition at $5K$. Such a value fits fairly well with the doping
phase diagram of cuprate  superconductors.

\end{abstract}


 \maketitle

\smallskip

\section{INTRODUCTION}
\smallskip
Quasi one dimensional two-leg ladders
Sr$_{14-x}$Ca$_{x}$Cu$_{24}$O$_{41}$ (Ca$x$) have been intensively
investigated over the last  years due to their fascinating
physical properties.\cite{Dagotto96,Dagotto99} Spin ladders
Sr$_{14-x}$Ca$_{x}$Cu$_{24}$O$_{41}$ have a spin-liquid ground
state with a finite spin-gap\cite{Dagotto92} and antiferromagnetic
(AF) long-range order\cite{Nagata99,Ohsugi99} coexisting with spin
singlets and hole pairs at low $T$. Furthermore, for samples with
the largest Ca concentration, an applied pressure in the range
30-80 kbar stabilizes a superconducting ground state in the ladder
planes with a transition temperature passing through a maximum at
10 K around 40 kbar. \cite{Nagata98} Consequently, clarifying the
interplay between spin and charge degrees of freedom in doped
antiferromagnets is a very important matter for understanding the
onset of a superconducting state in  spin ladders as well as in
high-$T_{c}$ superconductors. The
Sr$_{14-x}$Ca$_{x}$Cu$_{24}$O$_{41}$ system is intrinsically hole
doped as the stoichiometry implies an average copper valency of
2.25. The redistribution of the preexisting holes among  chains
and ladders in this composite  system, depending on the Ca content, applied
pressure and temperature is one of the most important factor which
controls the physical properties (in particular superconductivity) of these spin-ladders.

Optical measurements\cite{Osafune97}  and the calculation of the
Madelung potentials \cite{Mizuno97} have shown that for the $x=0$
compound  holes are staying essentially in the chains providing a
valency of 2.06 for the ladder Cu sites. Their localization in the
chains leads to an insulating  behavior. In addition, upon Ca
substitution holes are transferred from the chains to the ladders.
\cite{Kato96,Motoyama97,Osafune97} At the same time, the
longitudinal conductivity (along the c-axis) increases, leading to
quasi one dimensional (Q-1-D) metallic properties.
\cite{Motoyama97,Nagata98} This hole transfer can be caused by the
reduction of the distance between  chains and (Sr,Ca) layers,
\cite{McCarron88} which results in an enhancement of the
electrostatic potentials in the chains.\cite{Mizuno97}
Experimentally, however, the precise amount of  hole transfer is
still under discussion. Osafune \textit{et al}.\cite{Osafune97}
have found that the copper valency in the ladders increases from
$2.06$ up to $2.22$ ($n_{h}=0.22$ holes per Cu) when $x$ increases
from $0$ to $11$. The systematic measurements of the $^{63}$Cu
nuclear spin-spin relaxation time $T_{2G}$
(Ref.\onlinecite{Magishi98}) probing the spin correlation length
has shown that holes were doped into Cu$_{2}$O$_{3}$ ladder with a
content of $n_{h}\sim0.14$, 0.22 and 0.25 holes per Cu for $x$=6,
9 and 11.5, respectively. Recently, Isobe \textit{et
al}.\cite{Isobe00} have investigated the temperature dependence of
the crystal structure of Sr$_{0.4}$Ca$_{13.6}$Cu$_{24}$O$_{41}$
ladders by the Rietveld analysis of neutron diffraction data.
Using the bond-valence sum calculation they have estimated the
hole concentration in ladder planes as
$n_{ladder}(T=300K)\sim0.09$ per Cu1, \textit{i.e.} about three
times smaller than what was reported in Ref.
\onlinecite{Osafune97,Magishi98}.  Moreover, Isobe \textit{et
al}.\cite{Isobe00} found that holes located in the ladders tend to
move back into the chains at low temperature and that almost all
  holes were located in the chains near the liquid-He
temperature. A similar estimate for the hole content
$n_{ladder}(T=300K)\sim0.08$  for the Ca12 compound has been
recently deduced from polarization-dependent $X$-ray absorption
data by N\"{u}cker \textit{et al}.\cite{Nucker00} With a
decreasing $x$, $X$-ray data indicate only a marginal decrease of
$n_{ladder}(T=300K)\sim0.06$ for Ca0. More recently, Thurber
\textit{et al}.\cite{Thurber03} have reported NMR measurements of
the $^{17}$O nuclear quadrupole interactions $^{17}\nu_{Q}$ on the
Ca$x$ ($x$=0, 3, 8). They have related the changes in
$^{17}\nu_{Q}$ to the variation in the effective hole
concentration $\Delta n_{ladder}$ in the Cu$_{2}$O$_{3}$ ladder
layer and have found a
\begin{figure}[htbp]
\centerline{\includegraphics[width=0.85\hsize]{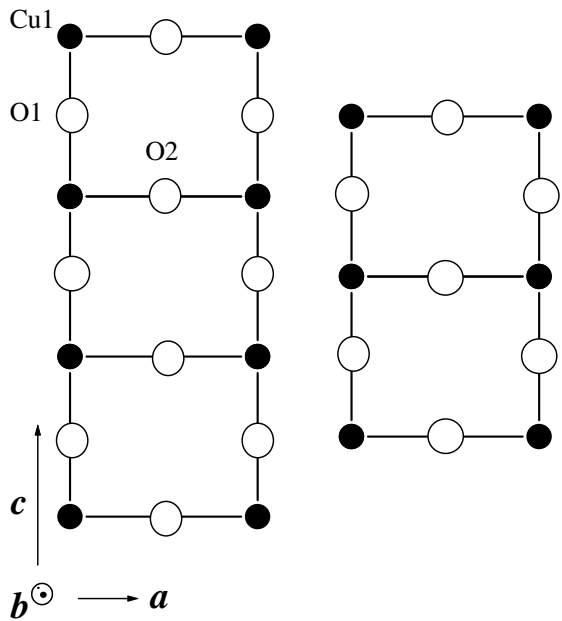}}
\caption{Schematic representation of the two-leg ladder layer.
Closed and open circles are Cu and O atoms, respectively.}
\label{struct}
\end{figure}
decrease of the hole number in Cu$_{2}$O$_{3}$ going to low temperature for all $x$ values
. In addition, they have noticed only a small
increase of the hole density of $\Delta n_{ladder}\sim0.04$ in the
ladders layer  going from Ca0 to Ca8.

The main effect of Ca doping on the structure of
Sr$_{14-x}$Ca$_{x}$Cu$_{24}$O$_{41}$ is a reduction of the
$\textbf{b}$ parameter. A similar conclusion is reached by
analyzing the effect of hydrostatic
pressure.\cite{Pachot99,PachotThese} It is the $\textbf{b}$
parameter which is most sensitive to pressure. Hence, we can
expect an increase of the ladder hole density  to take place as
pressure is increased. This assumption is in agreement with the
decrease of the electrical resistivity observed in Refs.
\onlinecite{Motoyama97,Isobe98,Nagata98} under high pressure.
Experimentally, however,   quantitative estimates for the hole
number transferred into the ladder subsystem under high pressure
are still missing.

In order to address such a question the present  work attempts to provide an  experimental determination of the hole
distribution among ladder and chain subsystems as a function of
temperature, doping, and pressure. The latter point is crucial
since pressure is a prerequisite for the stabilization of
superconductivity in Sr$_{14-x}$Ca$_{x}$Cu$_{24}$O$_{41}$.  We
present detailed copper and oxygen NMR studies of the nuclear
quadrupole interaction tensor, $^{17,63}\nu_{Q,\alpha}$, in Ca0
and Ca12 single crystal samples at ambient pressure and under the
pressure of 32 kbar which can drive the Ca12 system to the
situation where superconductivity is stabilized at low
temperature.  The measurements of $^{17,63}\nu_{Q,\alpha}$ were
performed at the different ion sites Cu1, O1 and O2 (see Fig. 1) in
the Cu$_{2}$O$_{3}$ ladder plane. The part concerning the ambient
pressure data is  consistent with the work of previous workers.
\cite{Thurber03} Our results emphasize an additional transfer of
holes from  chains to  ladders when pressure is applied in highly Ca-substituted samples.
\begin{figure}[htbp]
\centerline{\includegraphics[width=0.85\hsize]{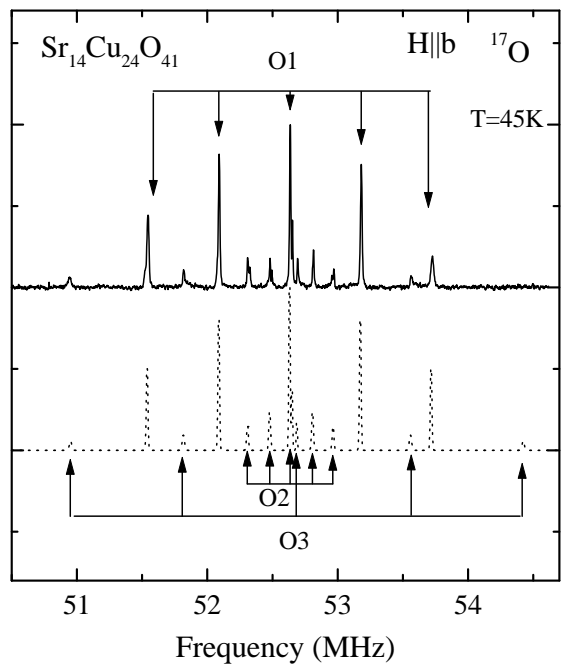}}
\caption{ $^{17}$O NMR spectrum in Sr$_{14}$Cu$_{24}$O$_{41}$ for
$H\parallel{\textbf{b}}$ measured at $T$=45 K under ambient
pressure (solid line) and the results its simulation (dotted
line).} \label{spectrum}
\end{figure}

\section{EXPERIMENT}

NMR studies were carried out on monocrystalline samples of
Sr$_{14}$Cu$_{24}$O$_{41}$ (Ca0) and
Sr$_{2}$Ca$_{12}$Cu$_{24}$O$_{41}$ (Ca12)  grown by the traveling
solvent floating zone method. \cite{Regnault99} The samples were
enriched with $^{17}$O isotope as described
elsewhere.\cite{Piskunov04} $^{63}$Cu($I=3/2$) and
$^{17}$O($I=5/2$) NMR measurements have been performed at a field
of $9.3$ Tesla with the usual Fourier transform method. The
components of the quadrupole interaction tensor,
$^{17,63}\nu_{Q,\alpha}, (\alpha=a, b, c)$  were determined from
$^{17}$O and $^{63}$Cu NMR spectra recorded for the different
orientations of the single crystals in the external magnetic field
using a simulation software taking into account quadrupolar
corrections to the Zeeman frequency in a second order perturbation
theory. A typical $^{17}$O NMR spectrum obtained for the Ca0
sample at ambient pressure (solid line) and the result of its
simulation (dotted line) are displayed in Fig. \ref{spectrum}. A
precise orientation of the single crystal in the magnetic field
along the corresponding axis has been reached through a fine
adjustment of the angular position of the pressure cell in the
magnetic field. As far as the Cu nucleus is concerned we used  the
second order shift of the central line as explained below but we
also checked on the sample Ca0 that the value of $^{63}\nu_{Q,b}$
thus obtained was in very good agreement with the first order
quadrupolar shift derived from the  frequency difference between
the central and the satellite lines.

NMR frequencies  up to the second order quadrupolar contribution
are given by the equations,
\begin{equation}
^{63}\nu _{\left( 1/2,-1/2\right) ,b}=(1+K_{b})\nu
_{0}+\frac{\left( \nu _{c}-\nu _{a}\right) ^{2}}{12\nu _{0}\left(
1+K_{b}\right) }\label{nuQ}
\end{equation}
for $H\parallel b$-axis where $K_{b}$, $\nu _{0}$, $\nu _{\alpha}$
are the Knight shift, the Larmor frequency in a diamagnetic
substance and the quadrupolar tensor components, respectively. For
$H\parallel c$-axis, $^{63}\nu _{\left( 1/2,-1/2\right) ,c}$ can
be expressed as,
\begin{equation}
^{63}\nu _{\left( 1/2,-1/2\right) ,c}=(1+K_{c})\nu
_{0}+\frac{\left( 3\nu_{b}-\left| \nu _{c}-\nu _{a}\right| \right)
^{2}}{48\nu _{0}\left( 1+K_{c}\right) }
\end{equation}
When the magnetic field is aligned along the $c$-axis, the quadrupolar
contribution to the NMR shift is about ten times larger than for the
case $ H\parallel b$ according to equations 1 and 2
and consequently must  be treated on
equal footings with the magnetic shifts.
\begin{figure}[htbp]
\centerline{\includegraphics[width=1\hsize]{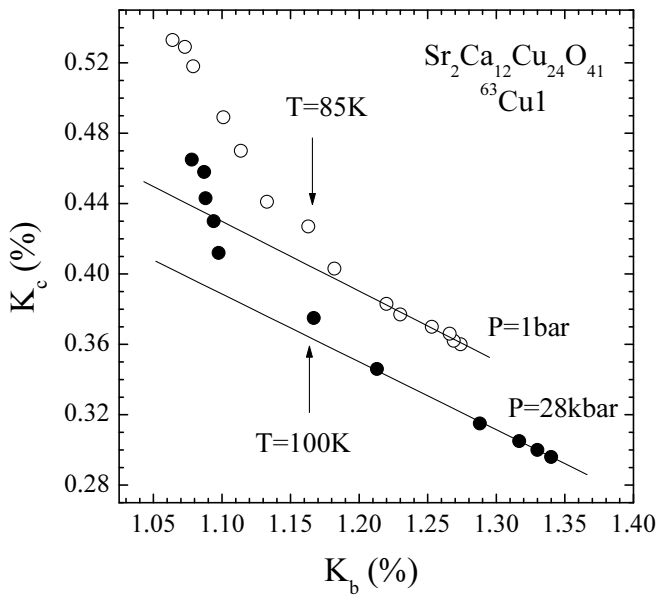}}
\caption{ $^{63}$Cu1 $K_{c}$ versus $K_{b}$ plots with temperature as the
implicit parameter at $P=1$bar and 28kbar in Ca12. Straight lines concern the low temperature domain.}
\label{fig10p1.eps}
\end{figure}

We have determined
$K_{c}$ (Eq. (2)) using the values $\nu _{a}=-4.2MHz$, $\nu
_{b}=15.7MHz$, $\nu _{c}=-11.5MHz$ estimated by Magishi {\it et-al}.
\cite{Magishi98} for the neighboring composition
$Sr_{2.5}Ca_{11.5}Cu_{24}O_{41}$ at low $T$. Assuming that $
K_{c,s}(P=1bar,T=0)=0$ in the spin gap regime at ambient pressure we
obtain:
$K_{c,orb}=K_{c}(T=0)=0.30\%$. Since the spin parts $K_{c,s}$ and $K_{b,s}$
are both proportional to the spin susceptibility $\chi _{s}(T)$ and given $%
K_{orb}$ which is temperature independent, the plot $K_{c}$ versus $K_{b}$
with temperature as the implicit parameter must be linear,
\begin{equation}
K_{c}(T)=\frac{A_{c}(0)}{A_{b}(0)}K_{b}(T)+\left[ K_{c,orb}-\frac{A_{c}(0)}{%
A_{b}(0)}K_{b,orb}\right] ,
\end{equation}
where $A_{c,b}$ are the uniform hyperfine fields.

$K_{c}$ versus $K_{b}$ plots at 1bar and 28kbar are shown in
figure. \ref{fig10p1.eps}. As seen in the figure, these plots are
not fitted by a straight line over the whole $T$ range. We can
ascribe the deviation of the data from straight lines to the
temperature dependence of $^{63}\nu _{Q}\equiv \left| \nu
_{b}\right| $, namely, it is due to $\nu _{Q}$ increasing  with
the temperature. Such an increase of $\nu _{Q}$ was reported in
several NMR studies of $\mathrm{Sr_{2}Ca_{12}Cu_{24}O_{41}}$
\cite{Carretta97,Takigawa98,Imai98}. In those works, it was
indicated that the variation of $\nu _{Q}$ is negligible in the
low $T$ regime but becomes quite appreciable above $T\approx
100K$. Therefore, $K_{c}(T)$ versus $K_{b}(T)$ plots were fitted
by straight lines in the temperature range below $100K$. As a
result of the fit, $\frac{ A_{b}(0)}{A_{c}(0)}=2.6\pm 0.1$ is
independent of the applied pressure. This finding agrees well with
$\frac{A_{b}(0)}{A_{c}(0)}=2.5$ derived in reference
\cite{Magishi98} . Using the ratio $\frac{A_{b}(0)}{A_{c}(0)}=2.6$
and $ K_{b,s}$ data $vs$ temperature one can derive with a little
bit of algebra the $T$ dependence of $\nu _{Q}$  displayed  on
figure.\ref{Cu1nQvsT} .

\section{ RESULTS}

\begin{figure}[htbp]
\centerline{\includegraphics[width=0.85\hsize]{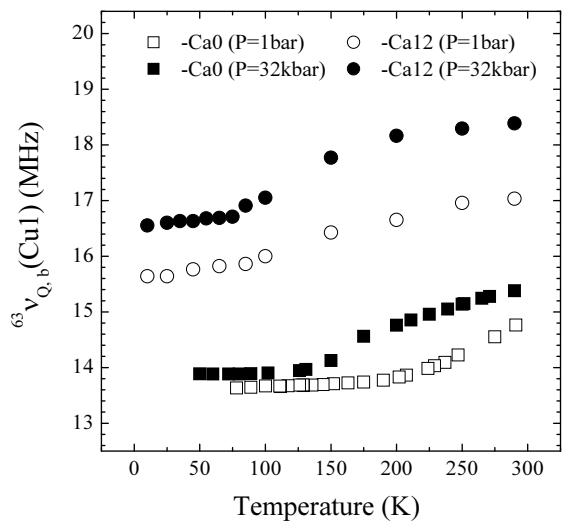}}
\caption{Temperature dependence of the nuclear quadrupole
frequency $^{63}\nu _{Q,b}$ at the ladder Cu1 sites in Ca0 and
Ca12 single crystals under ambient and high pressures.}
\label{Cu1nQvsT}
\end{figure}

\begin{figure}[htbp]
\centerline{\includegraphics[width=0.85\hsize]{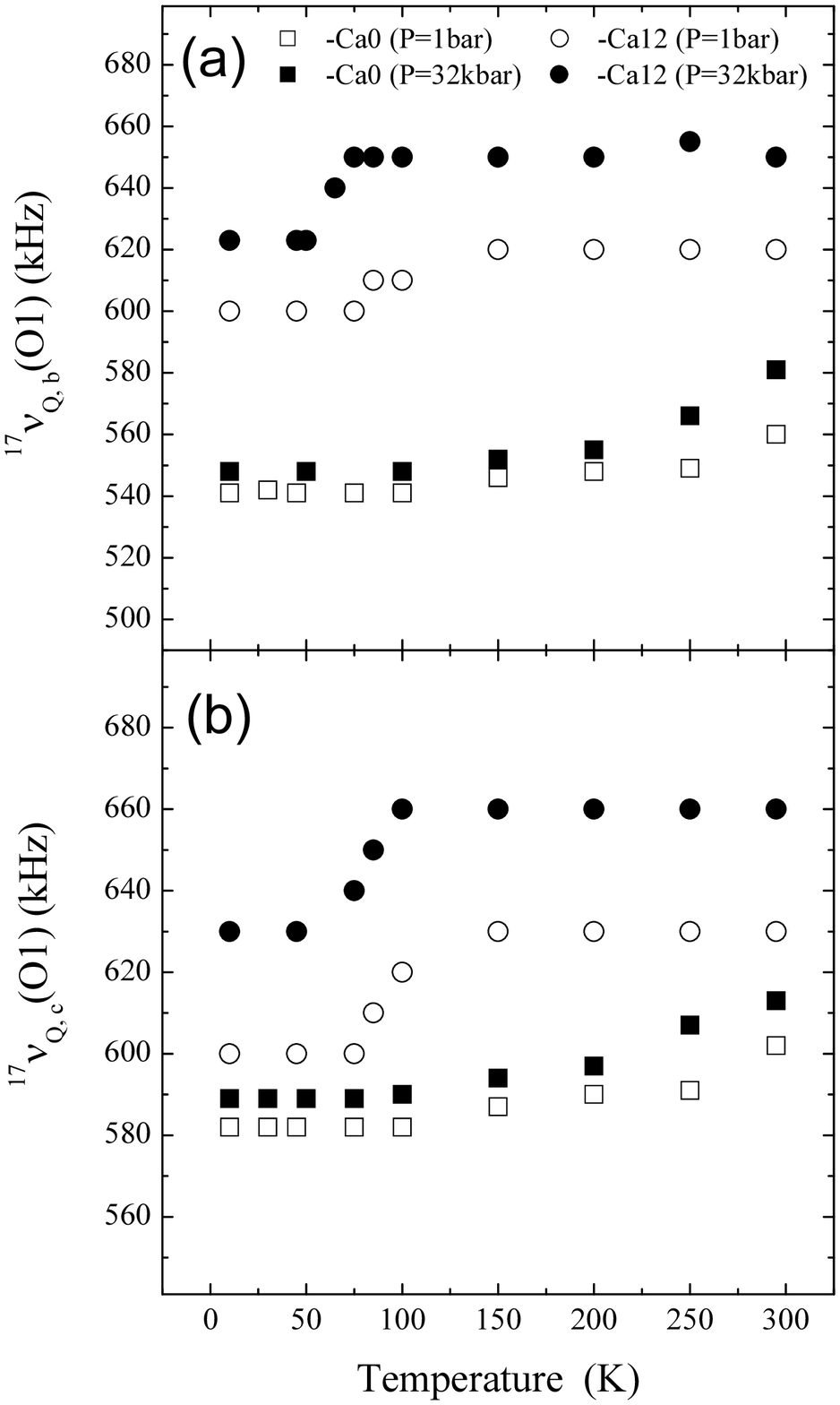}}
\caption{Temperature dependence of  $^{17}\nu _{Q,b(c)}$ at the
ladder leg O1 site in Ca0 and Ca12 for H$\|\textbf{b}$ (a) and for
H$\|\textbf{c}$  (b) under ambient and high pressures.}
\label{o1nQvsT}
\end{figure}

\begin{figure}[htbp]
\centerline{\includegraphics[width=0.85\hsize]{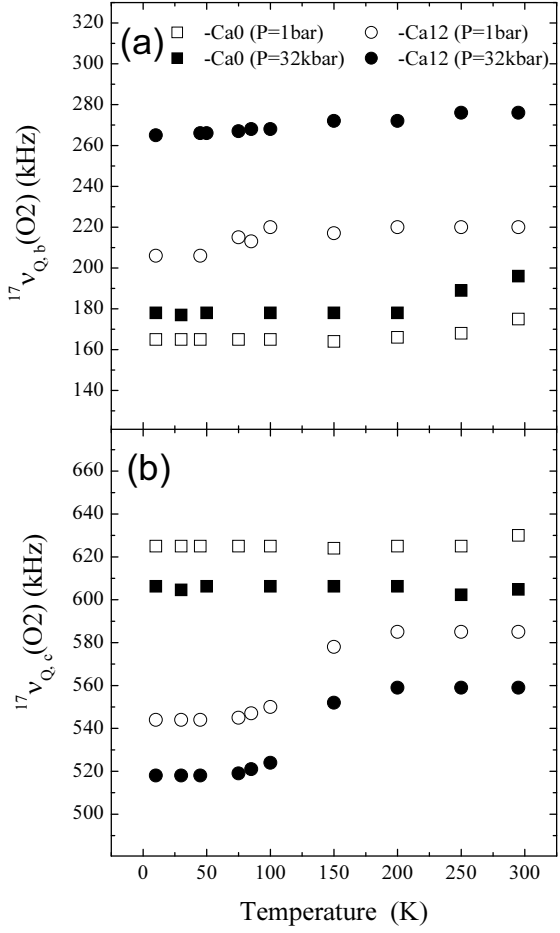}}
\caption{Temperature dependence of  $^{17}\nu _{Q,b(c)}$ at the
ladder rung O2 site in Ca0 and Ca12 for H$\|\textbf{b}$ (a) and
for H$\|\textbf{c}$  (b) under ambient and high pressures.}
\label{o2nQvsT}
\end{figure}

Temperature dependences of the $b$- and $c$-axis components of the
oxygen and copper nuclear quadrupole frequences, $^{17, 63}\nu
_{Q,b(c)}$, measured for the Cu1, O1 and O2 ladder sites in Ca0
and Ca12 samples under ambient and high pressures are presented in
Figures \ref{Cu1nQvsT}, \ref{o1nQvsT} and \ref{o2nQvsT}. For the Ca-
free  sample, $^{17, 63}\nu _{Q, \alpha}$, for the Cu1, O1, and
O2 sites depends on temperature only weakly below 150 K, however,
increases steeply above 150 K. As to the Ca12 compound, $\nu _{Q,
\alpha}$(Cu1,O1,O2) shows a dramatic change in the temperature range
$\Delta T=50\div200$ K and a $T$ independent behavior outside this
region.
%%%%%%%%%%%%%%%%%%%%%%%%%%%%%%%%%%%%%%%%%%%%%%%
As is seen in the figures, the quadrupole frequencies $\nu _{Q,
b}$(Cu1,O1,O2)  and $\nu _{Q, c}$(O1) reveal an increase
with  both the Ca content and an applied pressure while on the contrary
$\nu _{Q, c}$(O2) is  reduced. This fact is
indicative of an increase of the asymmetry parameter for EFG
tensor at the O2 site, $\eta$(O2), with increasing Ca content and
pressure.

Let us now consider  possible reasons for the evolution of
$^{17,63}\nu _{Q, \alpha} (\alpha=a, b, c)$ varying doping,
pressure and temperature.

In a semiempirical approach the
quadrupole frequency can be written as a sum of two
contributions,\cite{Cohen57}
\begin{equation}
\nu _{Q, \alpha}=(1-\gamma _{\infty })\nu _{l, \alpha}+\nu _{ h,
\alpha}, \label{lathole}
\end{equation}
where the first term describes the contribution from neighboring
ions in a point-charge model enhanced by the Sternheimer
antishielding factor $\gamma _{\infty }$ and $\nu _{h, \alpha}$ is
the term arising from holes in the orbitals of the ion itself. For
the Cu 3d and O 2p unfilled shells the hole contributions
$^{63}\nu_{h}$ and $^{17}\nu_{h}$ can be expressed as
follows\cite{Hanzawa93}
\begin{equation}
^{63}\nu_{h}=\frac{1}{2}\frac{^{63}Qe^{2}}{h}\frac{4}{7}\left\langle
r^{-3}\right\rangle_{3d}n_{3d}=\nu_{3d,0}n_{3d} \label{Cuhole}
\end{equation}

\begin{eqnarray}
^{17}\nu_{h,\alpha}=\frac{3}{20}\frac{^{17}Qe^{2}}{h}\frac{4}{5}\left\langle
r^{-3}\right\rangle_{2p}\left[n_{2p, \alpha}-\frac{n_{2p,
\beta}}{2}-\frac{n_{2p, \gamma}}{2}\right]\nonumber\\=
\nu_{2p,0}\left[n_{2p, \alpha}-\frac{n_{2p,
\beta}}{2}-\frac{n_{2p, \gamma}}{2}\right] (\alpha, \beta, \gamma
= a, b, c).\quad \label{Oxhole}
\end{eqnarray}

Here, $\nu _{3d,0}$ ($\nu_{2p,0}$) is the quadrupole frequency
when there is one hole in the 3d (2p) orbital and $n_{3d}$
($n_{2p, \alpha}$) is the hole number in the Cu 3d (O 2p,$\alpha$)
shell. If we take $^{63}Q=-0.211\times 10^{-24}$ cm$^{2}$,
$^{17}Q=-0.026\times 10^{-24}$ cm$^{2}$, $\left\langle
r^{-3}\right\rangle_{3d}=6.04$ a.u.\cite{Shimizu93}, $\left\langle
r^{-3}\right\rangle_{2p}=3.27$ a.u.\cite{Clementi64,Eremin97} we
get, $\nu_{3d,0}=77$ MHz, $\nu_{2p,0}=2.4$ MHz.

Quite generally, the lattice contribution to the quadrupole
frequency reads,
\begin{equation}
(1-\gamma _{\infty })\nu _{l, \alpha}=(1-\gamma _{\infty })\frac{3
\enskip
Qe}{2I(2I-1)h}\sum_{i}\frac{\partial^{2}}{\partial\alpha^{2}}\frac{e_{i}}{r_{i}},
\label{CuOxlat}
\end{equation}
where $e_{i}$ is the effective charge of the $i$-th ion and
$r_{i}$ is its position. The sum goes over all crystal sites
except for the one considered. In our calculations we used the
same values for $^{17}\gamma_{\infty}=-9$ and
$^{63}\gamma_{\infty}=-20$ as were found in high-$T_{c}$
cuprates.\cite{Shimizu93,Takigawa89} Then, for $\alpha$-component
of the quadrupole frequency at the oxygen $k$ sites ($k$=O1,O2),
$\nu_{Q, \alpha}(k)$ reads (in MHz),
\begin{eqnarray}
^{17}\nu_{Q, \alpha}(k) =2.4\left[n_{2p, \alpha}^{k}-\frac{n_{2p,
\beta}^{k}}{2}-\frac{n_{2p,
\gamma}^{k}}{2}\right]+\nonumber\\2.80\times 10^{-9}
\sum_{j}L_{\alpha}^{k}(j)e_{j}. \label{Oxholelat}
\end{eqnarray}
Here,
$L_{\alpha}^{k}(j)=\sum_{i}^{(j)}\frac{\partial^{2}}{\partial\alpha^{2}}\frac{1}{r_{i}}$
is the contribution of the ions located at one of the position $j$
($j$ = O1, O2, O3, Cu1, Cu2, Sr/Ca sites) to the
$\alpha$-component of the lattice sum for the $k$-th site ($k$=O1,
O2). Cu2 and O3 are copper and oxygen sites in the CuO$_{2}$
chains.

Similarly, for $\nu_{Q, \alpha}$(Cu1) we get,
\begin{equation}
\nu_{Q,\alpha}(Cu1)=77n_{3d}+1.66\times10^{-7}\sum_{j}L_{\alpha}^{Cu1}(j)e_{j}
\label{Cuholelat}
\end{equation}
It is important to notice that there is a rather large uncertainty
in the lattice contribution (Eq.(7)) to the quadrupole frequency
calculated by using a point charge model, which is an
oversimplified approximation. Furthermore, the Sternheimer factors
$^{63}\gamma_{\infty}$ and $^{17}\gamma_{\infty}$ are only crudely
determined in the ladders. This uncertainty conveys an important
uncertainty for the actual determination of the absolute hole
concentration in the ladder Cu$_{2}$O$_{3}$ layers. Fortunately,
the change in the lattice contribution caused by a variation of
the temperature, Ca doping and pressure are negligible. Hence, the
modification of the quadrupole frequency $\Delta\nu _{Q,
\alpha}=(1-\gamma _{\infty })\Delta\nu _{l, \alpha}+\Delta\nu _{
h, \alpha}$ observed in the experiment is dominated by the second
term $\Delta\nu _{ h, \alpha}$ in Eq. (4). In this case, the
uncertainty in the lattice contribution $(1-\gamma _{\infty
})\Delta\nu _{l, \alpha}$ can be neglected since it is
 a  negligible contribution altogether.

The change of quadrupole frequencies
$^{17}\nu_{Q, \alpha}$ and $^{63}\nu_{Q, \alpha}$ originating from
 the  hole transfer between   chains and  ladders can be expressed
as follows,
\begin{eqnarray}
\Delta\nu_{Q, \alpha}(k) =2.4\left[\Delta n_{2p,
\alpha}^{k}-\frac{\Delta n_{2p, \beta}^{k}}{2}-\frac{\Delta n_{2p,
\gamma}^{k}}{2}\right]+\nonumber\\2.8\times 10^{-9}
\sum_{j}L_{\alpha}^{k}(j)\Delta n_{j}+2.8\times 10^{-9}
\sum_{j}\Delta L_{\alpha}^{k}(j) e_{j}. \label{Chox}
\end{eqnarray}
\begin{eqnarray}
\Delta\nu_{Q,\alpha}(Cu1)=77\Delta
n_{3d}+1.66\times10^{-7}\sum_{j}L_{\alpha}^{Cu1}(j)\Delta
n_{j}\nonumber\\+1.66\times10^{-7}\sum_{j}\Delta
L_{\alpha}^{Cu1}(j)e_{j},\qquad \label{Chcu}
\end{eqnarray}
where $\Delta n_{3d}$ ($\Delta n_{2p, \alpha}^{k}$) is the
variation of  the hole number in the Cu 3d (O 2p,$\alpha$)
orbitals and $\Delta n_{j}$ is the change of the hole density in
the entire O 2p (or Cu 3d) orbital of the ion at the site $j$. The
third term in Eqs.~(\ref{Chox}) and (\ref{Chcu}) takes into
account a change of the lattice contribution due to a variation of
lattice parameters as a function of  temperature, pressure and
doping. These small corrections have been calculated for each
cases of the  $\nu_{Q}$ variation  using the results of Refs.
\onlinecite{Pachot99,PachotThese}. Next, we have assumed that the
holes occupy only the O $2p_{\sigma}$ orbitals, \textit{i. e.}
$n_{2p, b}^{k}=0$. Hence,
\begin{equation}
\Delta n(O1 (O2))= \Delta n_{2p, a}(O1 (O2))+\Delta n_{2p, c}(O1
(O2)).\label{oxdn}
\end{equation}
In addition we have assumed that the  holes transferred into the ladders  are
removed evenly from all atoms of the chains, \textit{i. e}. $\Delta
n$(O3)$=\Delta n$(Cu2)$\equiv\Delta n_{chain}$, leading to,
\begin{equation}
\Delta n_{chain}=\frac{14}{30}\Delta n(Cu1)+\frac{14}{30} \Delta
n(O1)+\frac{7}{30}\Delta n(O2). \label{chdn}
\end{equation}
This latter assumption has practically  no effect on  the final
result. For instance, the result of our calculations is hardly
modified if we assume that all  holes in the chains are located
either in  oxygen O 2p or copper Cu 3d orbitals. We have also
ignored any effect of covalent bonding between copper and ligand
oxygens. Since the transferred hyperfine fields at the O1 and O2
sites are not affected by pressure, doping and
temperature\cite{Piskunov04} a variation of $\nu_{Q}$ due to the
change of covalent effects exceeding 1-2 \% change in $\nu_{Q}$
seems very unlikely. The solution of the system of equations
Eqs.~(\ref{Chox}) - (\ref{chdn}) allows us to derive the change in
the hole concentration, $\Delta n$, for each orbitals.

\begin{figure}[htbp]
\centerline{\includegraphics[width=1\hsize]{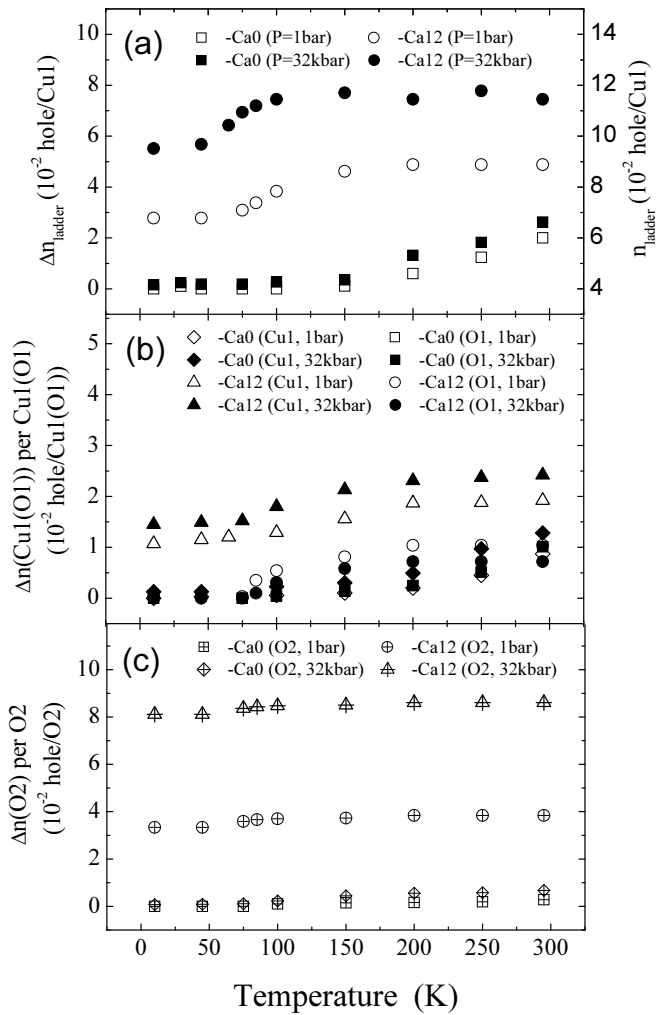}}
\caption{The hole distribution in Ca0 and Ca12 ladders under
ambient and high pressures. (a) $T$ dependence of the  hole
number per ladder Cu in the Cu$_{2}$O$_{3}$ ladder transferred
from the CuO$_{2}$ chains (left vertical axis) and of the total hole concentration
after calibration with the NEXAFS data, \textit{see text}
(right vertical axis). The distribution of transferred holes among the Cu1, O1 (b) and O2 (c) sites. }
\label{orbhvsT}
\end{figure}

\begin{figure}[htbp]
\centerline{\includegraphics[width=0.85\hsize]{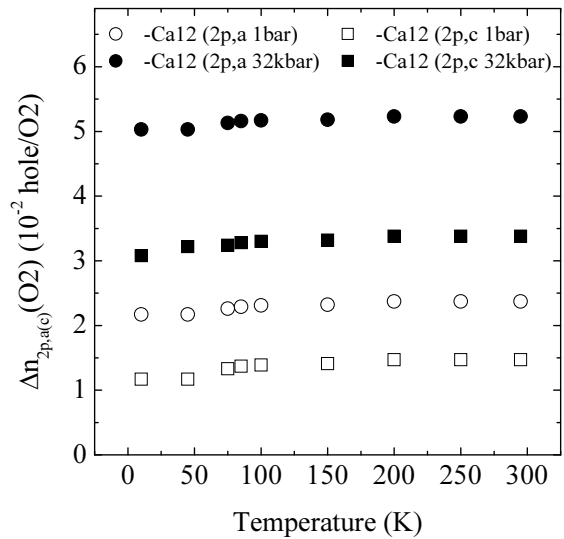}}
\caption{The distribution of holes located in the 2p orbitals of
the oxygen rung site O2 in Ca12 between the 2p,a and 2p,c
orbitals.} \label{holevsT}
\end{figure}

  The hole distribution in Ca0 and Ca12 ladder compounds under ambient and high pressures is summarized in Figs.
\ref{orbhvsT} and \ref{holevsT}.
  Fig. \ref{orbhvsT}(a) shows for  compositions Ca0 and Ca12 at ambient and high
pressures the temperature dependences of the  total hole number in the ladders (counted per
Cu1) which are transferred from the CuO$_{2}$ chains (left
vertical axis). This is the only  quantity our measurements can
determine accurately. In Fig.  \ref{orbhvsT}(a) the number of
holes $\Delta n_{ladder}$  transferred into the ladder plane of
Sr$_{14}$Cu$_{24}$O$_{41}$  was assumed arbitrarily  to be zero at
low $T$.
 The overall tendency for the hole density in the ladders is a decrease at low
temperature. Such a decrease
  can happen for the following reason: the backtransfer of holes
from  ladders to  chains. In this respect, we could ascribe the temperature of $\sim 200K$ below which on fig.\ref{o2nQvsT}a
$^{17}\nu_{Q,b}(O2)$ becomes temperature independent to the onset of a CDW state\cite{Gorshunov02} (or a hole
crystallization\cite{Abbamonte04}) in the ladders of Ca0. However, such an effect is not so clearly observed in the hole number on
the Cu1 sites. In addition we can rule out such an interpretation for the sudden drop of the hole number in Ca12 around $75 K$
since for this compound the CDW ground state is no longer stable at low temperature\cite{Vuletic03}.
The backtransfer
of  charges from the ladders to the chains as  temperature is
lowered, when the distance between the chains and the ladders is
shortened seems to contradict  the fact that a shrinkage of the
intersheet CuO$_{2}$-Cu$_{2}$O$_{3}$ distance leads to a hole
transfer towards the ladders. This discrepancy has been recently
resolved by Isobe \textit{et al}. \cite{Isobe00} These authors
have shown that it is the minimum O3-Cu1 distance of the modulated
structure which governs the hole density in the ladders. Moreover,
in spite of the average O3-Cu1 distance becoming shorter at low
temperature the minimum  does  expands. This expansion leads in
turn to a reverse hole transfer from ladders to  chains.

Using the bond-valence sum calculation, Isobe \textit{et
al}.\cite{Isobe00} have obtained that the  ladder layer of the Ca13.6
compound contains  1.3 holes per formula unit (\textit{f.u}.) (\textit{i.e.} 0.092 hole per Cu1)
at room temperature. In addition, according to this calculation, as the temperature is lowered
down to 5 K nearly all holes ($\sim$1 hole per \textit{f.u}. or 0.07 hole per Cu1) move
back to the chains. We note that our results give for Ca12 a much smaller
fraction (0.28 holes per \textit{f.u}. or 0.02 hole per Cu1) of the holes moving back into the
chains with decreasing temperature. Furthermore, since we only see
in the NMR/NQR spectrum the contribution from evenly distributed holes\cite{NQR}, our results
indicate that a rather substantial number of holes in Ca12 is
remaining on the ladders (possibly delocalized) even down to the lowest temperature.
Supporting the possibility that not all  holes in the Ca-rich
samples are backtransferred from the ladders to the chains at low
$T$ is the observation of an antiferromagnetic long-range ordering
taking place in the chains of  Ca-rich ($x\geq10$) spin ladders.
In the case when   six holes per \textit{f.u}. are located in
chains,   spin-singlet dimers form  and antiferromagnetic long-range order  does not
take place in chains. It is  the situation which prevails in Ca-poor
compounds with $x<10$ (Ref. \onlinecite{Nagata99}). The \textbf{presence}
of AF order in the chains of the
 Ca-rich ladders can serve as an indication that the hole number
in the chains is less than 6 holes per \textit{f.u}. In that event, "lone"
spins and/or "spin-trimers" can form in the chains inducing
staggered spin modulations on the remaining spin singlets and AF
long-range order in the spin liquid state is established.
\section{Discussion}
We have obtained that an increase of the Ca content up to 12 atoms
per \textit{f.u}. leads to only a small increase of the hole
density in the ladder layer ($\sim0.03$ hole/Cu1). This value is
much smaller than the  0.16 hole/Cu1 found in optical conductivity
measurements of Osafune \textit{et al.} for Ca12\cite{Osafune97}
but is quite close to the value 0.02 hole/Cu1 derived by in NEXAFS
measurements of N\"{u}cker \textit{et al}.\cite{Nucker00}
Interestingly, although the results of N\"{u}cker \textit{et al}.
and those of Osafune \textit{et al}. relative to the hole number
in Cu$_{2}$O$_{3}$ planes of Ca-rich samples are quite different
both investigations report the same absolute value
$n_{ladder}(300K)\approx0.06$ hole/Cu1 for the Ca0 compound at
room temperature. Consequently, if the hole density in the ladder of Ca0 is
really 0.06 holes per Cu1 at  room temperature then our results
imply that 0.04 hole/Cu1 still persist in the ladder plane of the
Ca-free sample at low $T$. Hence, it is necessary to add 0.04
hole/Cu1 to the amount $\Delta n_{ladder}$ of holes  transferred to
the ladders, see fig. \ref{orbhvsT}(a) (left axis), in order to obtain the
total hole number, $n_{ladder}$, located  in ladder layers as shown
on the right axis of the figure. Then, $n_{ladder}$(Ca12, 300K)
becomes 0.09 and 0.12 hole/Cu1 at 1 bar and 32 kbar respectively.

At this stage it is interesting to compare our findings for the
amount of holes sitting on the ladders in Ca0 with the
predictions derived from an \textit{ab initio} calculation by
Gell\'e and Lepetit.\cite{Gelle04} Following their calculation,
the number of holes on the ladders should be much
smaller than 1 per \textit{f.u}. in order to lead to a
localization of magnetic electrons in agreement with the
experimental observation of dimeric units separated by Zhang-Rice
singlets. According to fig. \ref{orbhvsT}(a) our experiment gives  0.04 per Cu1 or
(0.56 holes per \textit{f.u})  on the ladders at low
temperature. As far as the Ca12 compound is concerned, our
experiment gives under ambient pressure a number of 1 hole per \textit{ f.u.}
transferred into the ladders at low temperature. This figure
compares fairly well with the \textit{ab-initio} calculation made
for the Ca13.6 system \cite{Gelle04} predicting a small hole transfer
in-between 1 and 2 per \textit{ f.u.} in order to explain the
existence of an antiferromagnetic ground state at low temperature
in the heavily Ca-substituted samples.

An other important result of our study is the observation of an additional
 increase of $n_{ladder}$ in the ladder layer of the
Ca-rich compound under high pressure. The amount of increase is
comparable to what is  obtained  at ambient
pressure going from Ca0 to Ca12 . It is apparent, therefore, that the important role of
high pressure in achieving the conditions for the stabilization of
superconductivity in Ca12 is an increase of the hole doping of the
conducting Cu$_{2}$O$_{3}$ planes. On the other hand,  high
pressure leads to only a slight transfer of holes from chains to
ladders in Ca-free compound Ca0. Note, however, that the $\Delta
n_{ladder}$(Ca0, 32 kbar, 300K)$\simeq$0.0262 hole/Cu1, obtained in
our experiments, is quite close to the $\Delta n_{ladder}$(Ca3, 1
bar, 300K)$\simeq$0.0270 hole/Cu1 reported by Thurber \textit{et
al}.\cite{Thurber03} It is well known that the main effect of Ca
doping and of a hydrostatic pressure on the structure of
Sr$_{14-x}$Ca$_{x}$Cu$_{24}$O$_{41}$ is a reduction of the
\textbf{b} parameter.\cite{McCarron88,Isobe98,Pachot99} The
contraction of lattice parameters induced by a pressure of 30 kbar
corresponds to a substitution of three Sr by three Ca, \textit{i.
e.} $\textbf{b}(x=0, P=30 $kbar$)\approx\textbf{b}(x=3, P=1
$bar$)$.\cite{Pachot99} Thus, the equality between $\Delta
n_{ladder}$(Ca0, 32 kbar) and $\Delta n_{ladder}$(Ca3, 1 bar)
suggests that the transfer of holes from chains to ladders under
pressure and upon Ca doping is caused, at least for the Ca-poor
compounds, by the same reason namely,  a shrinkage of the
intersheet CuO$_{2}$-Cu$_{2}$O$_{3}$ distance.

The hole distribution among  Cu1, O1 and O2 sites of the ladders
as a function of temperature at different pressure and Ca content
is presented in Figs. \ref{orbhvsT}(b) and \ref{orbhvsT}(c). These
figures show that the holes transferred from the chains to
the ladders with increasing temperature sit mainly on the O1 and
Cu1 sites. Ca substitution leads to the transfer of holes which
reside on the Cu1 and  O2 rung sites. Finally, nearly all holes
transferred at room temperature into the ladders when pressure is applied  occupy the O2
sites.

Figure \ref{holevsT} shows the distribution of holes located in
the 2p orbitals of the oxygen rung site O2 in Ca12 between the
2p,a and 2p,c orbitals. As is seen in the figure, these holes go primarily in
the 2p orbitals oriented along the $a$-axis of the rungs at
increasing Ca content and  pressure. This is in
agreement with the conclusions of Ref. \onlinecite{Nucker00} where
an enhancement of the hole density in the O 2p,a orbitals in going
from Ca0 to Ca12 was found.

It has been shown in a number of theoretical studies that the spin
gap value $\Delta_{s}$ in the spectrum of spin excitations of
two-leg ladders is very sensitive to the degree of carrier doping  of the
Cu$_{2}$O$_{3}$ layer.\cite{Noack94,Venuti02} Namely, $\Delta_{s}$
 decreases as the hole number in the ladder plane is
increased. Our previous NMR
studies\cite{Mayaffre98,Piskunov00,Piskunov01} have revealed a
substantial reduction of the spin gap under pressure. Since the
applied pressure causes the transfer of holes from the chains to
the ladders it is reasonable to explain the spin gap reduction
 under pressure by an increase of the hole density in the
 ladder layer. As  mentioned above, the
ladder layers of Ca12 contain a finite amount of nonlocalized
holes even at low $T$ the number of which increasing under
pressure. Therefore, the growth of the residual spin
susceptibility at low $T$ at increasing pressure observed
in our previous works\cite{Mayaffre98,Piskunov00,Piskunov01} can be
related to  the increase  of single mobile holes in the
ladders at low $T$. Such a situation leads in turn to  a finite density of states at the Fermi
level which  contributes to the spin susceptibility as seen from the Korringa contribution to the NMR relaxation.

Note that the achievement of an appropriate carrier doping level
 due to the $b$-axis shrinkage under pressure is not a sufficient
condition for a stabilization of superconductivity in Ca$x$
compounds. Otherwise, even a less substituted Ca$x$ sample should
show superconductivity  applying a high pressure. Experimentally,
Sr$_{4}$Ca$_{10}$Cu$_{24}$O$_{41}$ did not reveal
superconductivity up to 85 kbar.\cite{Mori96}  Ca substitution has
an effect on the lattice  similar to high pressure although not
equivalent.\cite{Isobe98,Pachot99} The lattice parameter
$\textbf{a}$ in the Ca$x$ ladders shows hardly any change under
pressure but on the other hand is very sensitive to  Ca doping.
For instance, an hydrostatic compression of a Ca8 sample up to 110
kbar cannot reduce
 the $\textbf{a}$ parameter   down to its value in Ca12 under
30 kbar.\cite{Pachot99} Since a reduction of the $\textbf{a}$
parameter enhances the interladder interaction along the $a$-axis
it leads  to a more  two-dimensional ladder system. We suggest
that  prerequisites for the stabilization of superconductivity in
a spin-ladder are an adequate "two-dimensional degree" and a
large enough  carrier doping level. Ca substitution is essential
to achieve a  two dimensional character  while the necessary
carrier density  can only be
 attained  by pressure.

\section{ CONCLUSION}

In summary, we have reported a $^{63}$Cu and $^{17}$O NMR
investigation of the nuclear quadrupole interaction tensor,
$^{17,63}\nu_{Q,\alpha}$, in the hole doped spin ladder system
Sr$_{14-x}$Ca$_{x}$Cu$_{24}$O$_{41}$ ($x$=0 and 12) at ambient
pressure and under a pressure of 32 kbar when   Ca12 ladders are
superconducting. We have related the changes in
$^{17,63}\nu_{Q,\alpha}$ to a variation in the effective hole
concentration of the ladder layer and have
 determined the hole distribution among ladder and
chain systems as a function of temperature, Ca content and an
applied hydrostatic pressure. The  results are summarized
in Figs. \ref{orbhvsT} and \ref{holevsT}.

The overall tendency of the temperature dependence for the hole
density in the ladder layer is a reduction as  temperature is
lowered. The effective hole concentration in the Cu$_{2}$O$_{3}$
ladder plane of Ca0 compounds depends only
weakly on  the temperature  below 150 K but increases steeply above 150 K. The
additional hole number in ladder layers of Ca12 shows a dramatic
change in the temperature range $\Delta T=50\div200$ K and a
$T$-independent behavior outside this region. Holes transferred from
the  chains to the ladders at increasing temperature occupy
mainly the Cu $3d_{x^{2}-y^{2}}$ and the O 2p,c orbitals at the
Cu1 and O1 sites of ladders.

An increase of the Ca content and of the applied pressure leads to an
additional doping of the Cu$_{2}$O$_{3}$ ladder units. Ca
substitution leads to a transfer of holes in the orbitals of
Cu1 and rung oxygen O2 ions whereas under pressure nearly all
transferred holes occupy the O 2p,a orbitals of O2 oxygens.
A comparison between the effect of pressure and Ca substitution on the hole number
in the ladder layers and of the lattice parameters suggests that
the hole transfer from chains to ladders under pressure and upon
Ca doping is caused, at least for Ca poor compounds by a shrinkage
of the intersheet CuO$_{2}$-Cu$_{2}$O$_{3}$ distance. Furthermore, our results
indicate that a rather substantial number of holes in the ladder planes
of Ca12 is remaining delocalized down to the lowest temperature. We
suggest that the important role of high pressure for realizing
superconductivity in Ca12 is an increase of the hole density in the ladder
layers. However, the stabilization of a superconducting ground state
in the spin ladder Sr$_{14-x}$Ca$_{x}$Cu$_{24}$O$_{41}$ is
achievable only at a rather high "2D degree" in
the ladder plane. This two-dimensionality is obtained more by
 Ca substitution than by  high pressure.

Finally, it is worth putting the behavior of  superconducting spin
ladders in the context of  high $T_c$  (HTSC) superconductors.
Once enough 2-D character has been achieved by Ca doping,
superconductivity can then be stabilized at a  strongly pressure
dependent critical temperature. $T_c$ amounts to 5 K in Ca12 under
30 kbar\cite{Mayaffre98} and passes through a maximum at 10K in
Ca11.5\cite{Nagata97} and Ca13.6\cite{Takahashi97} under 45 and 35
kbar respectively. As suggested by the present NMR/NQR study,
pressure enhances the carrier concentration in the ladder planes. The
pressure (hole concentration) dependence of $T_c$ is also
reminiscent of the parabolic dependence found for $T_c$
\textit{versus} the carrier concentration in many HTSC
\cite{Tallon96}.

\begin{acknowledgments}

A.~Revcolevschi, U.~Ammerhal and G.~Dhalenne are gratefully
acknowledged for the samples preparation. We thank A.~Inyushkin
for helping us with the $^{17}O$ isotope exchange process. Y.P.
would like to thank S.~Verkhovskii for useful discussions.  Y.P.
acknowledges the CNRS and "Russian Science Support Foundation" for
a  financial support. The work was supported by the Ministry of
Sciences and Technology of the Russian Federation under contract
No 25/03 and the Russian Foundation for Basic Researches (Project
N 05-02-17846).
\end{acknowledgments}

\bibliographystyle{unsrt}

%\bibliography{LADDERS4}

\begin{thebibliography}{10}
\bibitem{NQR}
In case an entire hole were localized on a site it would shift the NMR signal away from the NMR frequency window.
\end{thebibliography}

\end{document}